\documentclass[preprint,review,12pt]{elsarticle}

\usepackage{amsmath}
\usepackage{amssymb}
\usepackage{amsthm}

\usepackage{lineno}

\usepackage{bm}
\usepackage{epstopdf}
\usepackage{hyperref}
\usepackage{booktabs}
\usepackage{textcomp}
\usepackage{xcolor}
\usepackage{setspace}
\usepackage{geometry}
\usepackage{adjustbox}
\usepackage{rotating}
\usepackage{subfigure}
\usepackage{appendix}
\biboptions{sort&compress}

\let\oldequation\equation
\let\oldendequation\endequation

\renewenvironment{equation}
  {\linenomathNonumbers\oldequation}
  {\oldendequation\endlinenomath}

\graphicspath{{Fig/}}

\newcommand{\bs}{\boldsymbol}
\newcommand\etal{\mbox{\textit{et al.}}}

\begin{document}

\begin{frontmatter}



\title{Modeling the boundary-layer flashback of premixed hydrogen-enriched swirling flames at high pressures}


\author[fir]{Shiming Zhang}
\author[fir]{Zhen Lu}
\ead{zhen.lu@pku.edu.cn}
\author[fir,sec]{Yue Yang}
\ead{yyg@pku.edu.cn}

\address[fir]{State Key Laboratory for Turbulence and Complex Systems, College of Engineering, Peking University, Beijing 100871, China}
\address[sec]{HEDPS-CAPT, Peking University, Beijing 100871, China}

\begin{abstract}
We model the boundary-layer flashback (BLF) of $\mathrm{CH_4}$/$\mathrm{H_2}$/air swirling flames via large-eddy simulations with the flame-surface-density method (LES-FSD), in particular, at high pressures.
A local displacement speed model tabulating the stretched flame speed is employed to account for the thermo-diffusive effects, flame surface curvature, and heat loss in LES-FSD.
The LES-FSD well captures the propagation characteristics during the BLF of swirling flames.
In the LES-FSD for lean $\mathrm{CH_4}$/$\mathrm{H_2}$/air flames at 2.5 bar, the critical equivalence ratio for flashback decreases with the increasing hydrogen volume fraction, consistent with the experiments.
This is due to the improved modeling of effects of the flame stretch and heat loss on the local displacement speed.
We also develop a simple model to predict the BLF limits of swirling flames.
The model estimates the critical bulk velocity for given reactants and swirl number, via the balance between the flame-induced pressure rise and adverse pressure for boundary-layer separation.
We validate the model against 14 datasets of $\mathrm{CH_4}$/$\mathrm{H_2}$/air swirling flame experiments, with the hydrogen volume fractions in fuel from 50\% to 100\%. The present model well estimates the flashback limits in various operating conditions.
\end{abstract}


\end{frontmatter}


\section{Introduction}\label{sec:intro}

Hydrogen is a promising fuel for zero-carbon, low-emission energy systems.
Meanwhile, burning hydrogen-enriched fuels increases the risk of flashback~\cite{Levinsky2021}.
Hydrogen-enriched flames may propagate upstream along the wall boundaries, i.e., boundary-layer flashback (BLF), due to the large flame speed and small quenching distance.
The flashback leads to either extinction or combustor damage.
It is one of the major threats to the safe and stable performance of combustors burning hydrogen-enriched fuels. 
To prevent the devastating events, especially for practical applications at high pressures, predicting the flashback limit is a key issue, which can be characterized as a minimum bulk velocity~\cite{Baumgartner2013} or a maximum equivalence ratio~\cite{Khateeb2021} for flame stabilization in lean premixed burners.

The BLF of non-swirling flames has been extensively studied~\cite{Lewis1943,Eichler2011,Gruber2012,Baumgartner2013,Baumgartner2015,Hoferichter2017,Vance2022,Goldmann2022}.
Lewis and von Elbe~\cite{Lewis1943} proposed the critical gradient model on the BLF limit.
They neglected the flame-flow interaction and suggested that BLF happens when the gradient of flame speed in the boundary layer exceeds the gradient of the flow velocity.
Recent experiments~\cite{Eichler2011} and direct numerical simulation (DNS)~\cite{Gruber2012} showed that the BLF of non-swirling flames propagates as small-scale bugles.
These bugles indicate that BLF initiates at the location of boundary-layer separation.
Accordingly, Hoferichter \etal~\cite{Hoferichter2017} modeled the BLF limit of premixed hydrogen-air flames in confined channels based on a boundary-layer separation criterion~\cite{Stratford1959} and a power-law scaling of the turbulent burning velocity.

The swirling flow is widely utilized to enhance mixing and flame holding~\cite{HuangY2009,Vignat2022}.
During the BLF in a swirl burner, the experiments~\cite{Karimi2015,Ebi2016,Ebi2018,Ebi2021} observed that a flame tongue, a convex-shaped large-scale flame front, propagates upstream along the central bluff body.
The flame tongue rotates along the bulk-flow direction for CH$_4$/air and CH$_4$/H$_2$/air flames at 1 bar~\cite{Ebi2016,Ebi2018}, whereas the flame front swirls against the bulk-flow direction for CH$_4$/H$_2$/air flames at 2.5 bar~\cite{Ebi2021}.
The mechanism for the switch of flame propagation modes is still unclear.


In particular, the swirling flames with higher hydrogen-enrichment levels are more prone to flash back even at a lower laminar flame speed.
To scale the BLF limits, Ebi \etal~\cite{Ebi2021} proposed a criterion based on a Karlovitz number defined with the flame extinction time scale and the shear rate in the boundary layer, but the model relying on the experimental measurement is not complete.
They concluded that both the models based on the critical gradient and the boundary-layer separation are hard to capture the BLF limits at different hydrogen-enrichment levels. 

The above studies showed that an accurate prediction of the BLF limit of swirling flames is challenging.
There are two possible reasons.
First, a direct application of the non-swirling models ignores the effect of the large-scale flame tongue and its propagation pathway~\cite{Ebi2016,Bailey2021}.
Second, the employed turbulent burning velocity models are not accurate enough for the wide range of conditions (e.g., pressures and fuels)~\cite{Ebi2021,Lu2022}.
Consequently, \textit{ad hoc} adjustments on model coefficients are required~\cite{Ebi2021}.

Besides experiments, the large-eddy simulation (LES) is a promising approach to gain insights into the BLF mechanism of swirling flames, regarding the reasonable resolution and affordable computational cost.
Several groups~\cite{Lietz2015,Jiang2021,Xia2022} have conducted LES for the BLF of swirling flames reported in the experiments at atmospheric pressure.
Lietz \etal~\cite{Lietz2015} captured the flame tongue structure in $\mathrm{CH_4}$/$\mathrm{H_2}$/air and $\mathrm{CH_4}$/air swirling flames at 1 bar using a flamelet model.
Jiang \etal~\cite{Jiang2021} employed the flamelet/progress variable and the flamelet generated manifold to model the BLF of swirling flames with fuel stratification and boundary heat loss.
Xia \etal~\cite{Xia2022} investigated the effects of the numerical boundary conditions on the BLF of CH$_4$/H$_2$/air swirling flames at 1 bar with the artificial thickened flame model.
They obtained a channel-like BLF for the swirling flame with an adiabatic central bluff body, and the large-scale flame tongue was observed for the non-adiabatic case with wall temperature of 350 K. 
On the other hand, there lacks a numerical study on the BLF process and the limit of swirling flames at elevated pressures.

The flame stretch effect is critical to modeling the turbulent flame propagation of hydrogen-enriched flames at high pressures.
Experiments~\cite{Venkateswaran2015,Abbasi-Atibeh_2019,Ahmed2021} and DNS~\cite{Lu2020,Rieth2022,Berger2022} investigated the mechanism  of the thermo-diffusive effects on the acceleration of lean hydrogen-enriched flames, where the super-adiabatic flame temperature, strongly wrinkled flame surface, and accelerated local propagation have been observed.
Thus, simulations of the hydrogen-enriched swirling flames require a proper model for the thermo-diffusive effects on flame propagation.

In the flame-surface-density (FSD) method, Zhang \etal~\cite{Zhang2021} developed a local displacement speed model to incorporate the thermo-diffusive effects via the stretch factor, which improves the LES-FSD result for turbulent premixed flames of lean H$_2$/air mixtures at high pressures.
However, this model does not consider the heat loss through the wall, and the heat loss is important in the LES-FSD for a flame propagating in a swirl burner with a central bluff body.
In addition, a model of the turbulent burning velocity for a wide range of conditions was developed~\cite{You2020,Lu2020,Lu2022}, in which the modeling of the stretch factor is crucial to characterize different fuels at high pressures.
Thus, an improved model of the stretch factor can facilitate predicting the BLF limit accurately, in particular, for fuel-lean mixtures at high pressures.

The objective of the present study is twofold: investigating the BLF process in the swirl burner with a central bluff body via LES-FSD, and developing a simple model predicting the BLF limit of swirling flames.
The rest of this paper is organized as follows.
The models in the LES-FSD method and the simulation setup are described in Sections~\ref{sec:methods} and \ref{sec:sim}, respectively.
The LES-FSD results are discussed in Section~\ref{sec:results}.
The BLF limit model for swirling flames is developed in Section~\ref{sec:FLmodel}.
Conclusions are drawn in Section~\ref{sec:conclusion}.

\section{LES-FSD method}\label{sec:methods}

\subsection{Transport equations}\label{sec:FSD}

In the LES-FSD for turbulent combustion, the filtered mass and momentum conservation equations were solved with four filtered scalars, the progress variable $\widetilde{c}$, generalized FSD $\Sigma\equiv\overline{\left\vert\nabla c\right\vert}$, mixture fraction $\widetilde{Z}$, and enthalpy $\widetilde{h}$,
where $\overline{q}$ and $\widetilde{q}$ denote the spatial and Favre filterings of a variable $q$, respectively.
The progress variable is defined as $c\equiv \left(T-T_{ub}\right) / \left( T_{b}-T_{ub} \right)$, where $T$ is the temperature, and subscripts $ub$ and $b$ denote the quantities in unburned reactants and burned products, respectively.
The enthalpy and mixture fraction are transported to account for the effect of heat loss and the variation of equivalence ratio.

The governing equations are
\begin{eqnarray}
	&&\frac{\partial\overline{\rho}}{\partial t}
	+ \nabla\cdot(\overline{\rho}\widetilde{\bs{u}})
	= 0,
	\label{eq:mass}
	\\
	&&\frac{\partial(\overline{\rho}\widetilde{\bs{u}}) }{\partial t}
	+ \nabla\cdot(\overline{\rho}\widetilde{\bs{u}}\widetilde{\bs{u}})
	+ \nabla\cdot(\overline{\rho}\widetilde{\bs{u}\bs{u}}-\overline{\rho}\widetilde{\bs{u}}\widetilde{\bs{u}})
	=
	- \nabla\overline{p}
	+ \nabla\cdot\overline{\bs{\tau}},
	\label{eq:momentum}
	\\
	&&\frac{\partial(\overline{\rho}\widetilde{c}) }{\partial t}
	+ \nabla\cdot(\overline{\rho}\widetilde{\bs{u}}\widetilde{c})
	+ \nabla\cdot(\overline{\rho}\widetilde{\bs{u}c}-\overline{\rho}\widetilde{\bs{u}}\widetilde{c})
	=
	\langle \rho s_d \rangle_A  \Sigma,
	\label{eq:c}
	\\
	&&\frac{\partial \Sigma}{\partial t}
	+ \nabla\cdot(\widetilde{\bs{u}}\Sigma)
	=
	- \nabla\cdot(\langle \bs{u} \rangle_A-\widetilde{\bs{u}})\Sigma
	+ \left(\nabla\cdot\widetilde{\bs{u}}-\bs{N}:\nabla\widetilde{\bs{u}} + \frac{ \Gamma \sqrt{k}}{\hat{\Delta}}\right)\Sigma
	\nonumber\\
	&&\qquad\qquad\qquad\qquad
	- \nabla\cdot \left( \langle s_d \rangle_A \langle \bs{n} \rangle_A\Sigma\right)
	+ \langle s_d \rangle_A \langle \kappa \rangle_A \Sigma
	- \frac{\alpha_N s_{L}^0(\widetilde{h},\widetilde{Z}) \Sigma^2}{1-\widetilde{c}},
	\label{eq:FSD}
	\\
	&&\frac{\partial(\bar{\rho}\tilde{h}) }{\partial t}
	+ \nabla\cdot(\overline{\rho}\widetilde{\bs{u}}\widetilde{h})
	+ \nabla\cdot(\overline{\rho}\widetilde{\bs{u}h}-\overline{\rho}\widetilde{\bs{u}}\widetilde{h})
	=
	\nabla\cdot\overline{(\rho D \nabla h) }
	+ {\overline{\mathcal{Q}}_{h}},
	\label{eq:enthalpy}
	\\
	&&\frac{\partial(\bar{\rho}\tilde{Z} ) }{\partial t}
	+ \nabla\cdot(\overline{\rho}\widetilde{\bs{u}}\widetilde{Z})
	+ \nabla\cdot(\overline{\rho}\widetilde{\bs{u}Z}-\overline{\rho}\widetilde{\bs{u}}\widetilde{Z})
	=
	\nabla\cdot\overline{(\rho D \nabla Z) },
	\label{eq:mixfrac}
\end{eqnarray}
where $t$, $\rho$, $p$, $\bs{u}$, and $\bs{\tau}$ are the time, density, pressure, velocity, and viscous stress, respectively; $\langle q \rangle_A \equiv \overline{q\left\vert\nabla c\right\vert}/ \overline{\left\vert\nabla c\right\vert}$ denotes the average of $q$ over the flame surface;
the surface-averaged mass flux is modeled as $ \langle \rho s_d \rangle_A = \overline\rho \langle s_d \rangle_A$~\cite{Hawkes2000b,Chakraborty2009};
$\langle \bs{n} \rangle_A = - \nabla \widetilde{c}/{\Sigma}$ is the modeled surface-averaged normal vector~\cite{Chakraborty2009};
$ \langle \kappa \rangle_A = \nabla\cdot \langle \bs{n} \rangle_A$ is the modeled surface-averaged curvature;
$\alpha_N=1-\langle \bs{n} \rangle_A \cdot \langle \bs{n} \rangle_A$ is an orientation factor~\cite{Hawkes2000b};
the tensor $\bs{N}=\langle \bs{n}\rangle_A\langle \bs{n} \rangle_A + (1-\langle \bs{n} \rangle_A \cdot \langle \bs{n} \rangle_A)\bs{I}/3$ models the strain rate~\cite{Hawkes2000a}, with the unit tensor $\bs{I}$;
$\hat{\Delta} = 5\Delta$ is the filter size~\cite{Boger1998} for scalars in combustion, where $\Delta$ is the filter size for the mass and momentum equations;
$\Gamma=0.75\exp\left[-1.2/(u'_{\Delta}/s_{L}^0(\widetilde{h},\widetilde{Z}))^{0.3}\right](\hat{\Delta}/\delta_{L}^0(\widetilde{h},\widetilde{Z}))^{2/3}$ is an efficiency function~\cite{Angelberger1998} with the subgrid velocity fluctuation $u'_{\Delta}=\sqrt{2k/3}$; $k$ is the subgrid turbulent kinetic energy.

The laminar flame speed $s_{L}^0(\widetilde{h},\widetilde{Z})$ and the flame thermal thickness $\delta_{L}^0(\widetilde{h},\widetilde{Z})$ were calculated from non-adiabatic one-dimensional freely propagating flames.
The non-adiabatic flame simulations are detailed later in the modeling of $\langle s_d\rangle_A$.
The heat transfer term $\overline{\mathcal{Q}}_{h} =  - h_w \bs{n}_w \cdot \nabla T $ accounts for the boundary heat loss,
where $h_w$ is the wall heat transfer coefficient, and $\bs{n}_w$ is the unit vector normal to the wall.

\subsection{Modeling of the local displacement speed}\label{sec:Sdmodel}

Modeling $\langle s_d \rangle_A$ plays an important role in FSD~\cite{Hawkes2000b,Zhang2021,Chakraborty2022,Yuvraj2022}.
The displacement speed $s_d$ is the propagation speed of a flame front relative to the convective flow.
The mass flux was often modeled as a constant as $\overline{\rho}\langle s_d\rangle_A = \rho_{ub} s_L^0$,
which is determined by an unstretched one-dimensional freely propagating flame~\cite{Boger1998,Chakraborty2009} with the laminar flame speed $s_L^0$, but recent studies showed that the flame stretch can significantly alter the local flame speed of lean hydrogen flames at elevated pressures due to the thermo-diffusive effects~\cite{Venkateswaran2015,Abbasi-Atibeh_2019,Lu2020,Rieth2022}.
Considering the flame stretch and curvature effects~\cite{Zhang2021}, we propose the model
\begin{equation}\label{eq:sdm2}
	\langle s_d \rangle_A
	=
	\dfrac{\rho_{ub} I_0 s_{L}^0 (\widetilde{h},\widetilde{Z})}{\overline\rho}
	-
	D{\langle \kappa \rangle_A}.
\end{equation}
Here, $D$ is the molecular diffusivity of the gas mixture, $I_0 = s_L( K, \widetilde{h}, \widetilde{Z} ) / s_L^0 ( \widetilde{h}, \widetilde{Z} )$ is the stretch factor,
where $s_{L}(K, \widetilde{h},\widetilde{Z})$ is the consumption speed of a stretched laminar flame at given Karlovitz factor $K$, enthalpy, and mixture fraction.
As the preferential diffusion of hydrogen is crucial to hydrogen-enriched flames, we model the thermo-diffusive effects via the stretch factor $I_0$.
It accounts for the variation of the local flame speed, thus influencing the transport of $\widetilde{c}$ and $\Sigma$ in Eqs.~\eqref{eq:c} and \eqref{eq:FSD}, respectively.
In LES-FSD simulations, $I_0$ is retrieved from a lookup table generated with separate laminar stretched flame simulations.
The Karlovitz factor $K$ in the flow field is modeled as
\begin{equation}\label{eq:K}
	K =
	\frac{ \delta_{L}^0(\widetilde{h},\widetilde{Z}) }
		 {s_{L}^0(\widetilde{h},\widetilde{Z})}
	\left(	\nabla\cdot\widetilde{\bs{u}}-\bs{N}:\nabla\widetilde{\bs{u}}  + \frac{\Gamma \sqrt{k}}{\hat{\Delta}}  \right).
\end{equation}

The displacement speed model in Eq.~\eqref{eq:sdm2} accounts for the flame stretch and curvature effects.
Validations~\cite{Zhang2021} confirmed that the model improves LES-FSD predictions for turbulent premixed flames of adiabatic lean hydrogen/air mixtures at high pressures.
In particular, the present model in Eq.~\eqref{eq:sdm2} further incorporates effects of the boundary heat loss and fuel stratification via additional dimensions of $\widetilde{h}$ and $\widetilde{Z}$.

To build the lookup table on $I_0$, we calculate the laminar counterflow flames and unstretched flames with different equivalence ratios and degrees of heat loss.
The effects of heat loss is considered in the laminar flame simulation via a modified energy equation~\cite{Proch2015}
\begin{equation}\label{eq:1Dflame}
    \dot{m} \nabla T
	=
	\nabla\cdot(\lambda\nabla T)
	- \sum_{i=1}^{n_s} c_{p,i} \bs{j}_{i}\cdot\nabla{T}
	- (1-f_l)\sum_{i=1}^{n_s} h_i \dot{\omega_i}W_i,
\end{equation}
where $\dot{m}$, $\lambda$, and $n_s$ are the mass flux rate, thermal conductivity, and number of species, respectively;
$c_{p,i}$, $\bm{j}_i$, $h_i$, $\dot{\omega}_i$, and $W_i$ denote the heat capacity, diffusive mass flux, enthalpy, reaction rate, and molecular weight of the $i$-th species, respectively.
The last term on the right-hand side of Eq.~\eqref{eq:1Dflame} is a scaled energy source term~\cite{Proch2015}, where $f_l$ is a heat loss factor.
For $f_l=0$, the energy equation is degenerated into the adiabatic one.
Increasing $f_l$ enhances heat loss and reduces $h$ on the burned side of flames.
Moreover, the flame solutions obtained with different $f_l$ are parameterized by $h$ on the burned side.

The laminar flames were calculated using Cantera~\cite{cantera} with a detailed chemical mechanism for 38 species and 291 reactions~\cite{FFCM-1}.
We started from the adiabatic simulation with $f_l=0$. The heat loss was gradually increased by raising $f_l$ up to 0.5 until the unstretched flame is quenched.
For each value of $f_l$, counterflow flames with different strain rates $a$ were calculated until extinction.
From solutions for the unstretched and stretched flames, we stored $I_0=s_L/s_L^0$, $s_L^0$, and $\delta_L^0$ against $K=a \delta_L^0 / s_L^0$, $h$, and $Z$.
In the following LES-FSD simulations, these values were retrieved from the table with presumed delta-distributions of $K$, $h$, and $Z$.

\section{Simulation overview}\label{sec:sim}

\subsection{Configurations}\label{sec:config}

\begin{figure}
	\centering
	\includegraphics[width=0.6\textwidth]{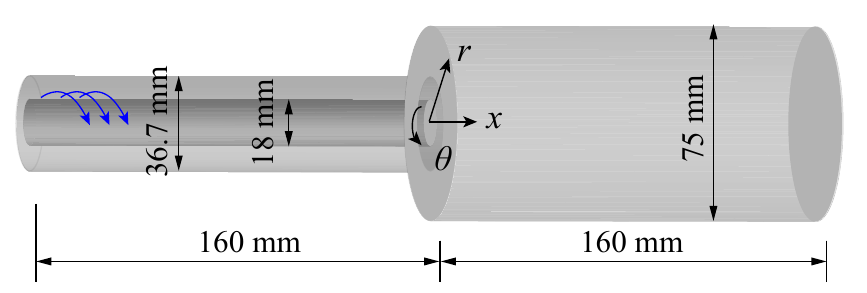}
	\caption{
        Schematic of the computational domain for LES-FSD, where the blue arrows denote the swirling inflow direction.
        }
	\label{fig:config}
\end{figure}

We simulate the flame stabilization and flashback to obtain the BLF limits of CH$_4$/H$_2$/air swirling flames with different hydrogen volume fractions $X_{\mathrm{H}_2}$.
The LES-FSD simulations correspond to the experiments with a swirl burner at $p=2.5$ bar in Ebi \etal~\cite{Ebi2021}.
As sketched in Fig.~\ref{fig:config}, the computational domain consists of a mixing tube and a combustion chamber.
The inner diameter, outer diameter, and length of the mixing tube are 18 mm, 36.7 mm, and 160 mm, respectively.
The combustion chamber has a diameter of 75 mm and a length of 160 mm.
The inlet and outlet are located at the bottom and the top of the computational domain, respectively.
A cylindrical coordinate was applied, with the origin at the center of the mixing tube outlet, axial $x$-direction, azimuthal $\theta$-direction, and radial $r$-direction.

In simulations, the fresh premixed gas was supplied through the inlet, with the preheat temperature $T_{ub}=473$ K, axial bulk velocity $U_x=15$ m/s, and swirl number $S=0.7$.
The BLF limit was determined as a critical equivalence ratio $\phi_{cr}$ for the occurrence of BLF by conducting a series of cases with a range of $\phi$.
As listed in Table~\ref{tab:sim_cases}, four $X_{\mathrm{H}_2}$ from 50\% to 80\% were adopted.
Cases A1 to A3 have the experimental operating conditions with $\phi_{cr}$ reported in experiments, and case E1 with $\phi=0.558$ has typical BLF of the swirling flame.

\begin{table} 	
    \centering
	\caption{Operating conditions in the LES-FSD for the BLF of CH$_4$/H$_2$/air swirling flames.}
	\setlength{\tabcolsep}{3mm}
	\begin{tabular}{ccccc}
		\hline
		$\mathrm{case}$ &$X_\mathrm{H_2}\;(\%)$ &$p\;(\mathrm{bar})$    &$T_{ub}\;(\mathrm{K})$    &$U_{x}\;(\mathrm{m/s})$    \\
		\hline
		A1	&50.0	&2.5	&473    &15 \\
		A2	&60.0	&2.5	&473	&15 \\	
		A3	&70.0	&2.5	&473	&15 \\
		E1  &80.0	&2.5	&473	&15 \\
		\hline
	\end{tabular}
	\label{tab:sim_cases}
\end{table}

To trigger BLF and determine the BLF limit, a lean stabilized flame was first simulated for each operating condition.
Then the equivalence ratio of the inlet stream was added by the increment $\Delta\phi=0.025$.
The simulation time for each equivalence ratio is at least 30 flow-through times of the mixing tube to observe whether BLF happens.
If BLF happens, we take the equivalence ratio of the stabilized case as the critical one $\phi_{cr}$.
Otherwise, the above procedure repeats until the BLF.
%
Note that the flame front may intermittently propagate into the mixing tube near flashback, which was observed in experiments~\cite{Schneider2020} and our simulations.
We determined the BLF state only when the flame front reaches the middle of the mixing tube at $x=-80$ mm.

\subsection{Numerical implementation}\label{sec:numerics}

We solve the governing equations for LES-FSD in Eqs.~\eqref{eq:mass} to \eqref{eq:mixfrac} using the NGA code~\cite{Desjardins2008}.
The momentum equations were discretized with a second-order, centered, kinetic-energy conservative scheme.
The third-order weighted essentially non-oscillatory scheme~\cite{Liu1994} were employed for convection terms in the scalar transport equations.
A semi-implicit Crank--Nicolson scheme~\cite{Pierce2001} was applied for the time marching of the transport equations.
The dynamic Smagorinsky model~\cite{Pierce2004} was employed to close the subgrid stresses, turbulent kinetic energy, and scalar fluxes.
More details on the LES-FSD implementation can be found in Ref.~\cite{Zhang2021}.

The computational domain was discretized by a mesh of 2 million cells.
The near-wall mesh was refined to ensure 15 grid points within $y^+ = 30$, where $y^+$ denotes the non-dimensional wall distance.
The mesh convergence test is given in~\ref{sec:atmospheric_pressure}.
The time stepping was set to ensure that the Courant--Friedrichs--Lewy number is less than 0.5.
We employed a separate simulation on a periodic mixing tube to generate the inlet velocity.
A linear forcing method~\cite{Carroll2013} was adopted to obtain a swirl number of 0.7, the same as that in the experiment.
Other boundaries were set as no-slip walls.
The temperature of the central bluff body was set to $T_{ub}$.

\section{LES-FSD of the BLF}\label{sec:results}

\subsection{BLF process of swirling flames}\label{sec:flameshape}

\begin{figure}
	\centering
	\includegraphics[width=\textwidth]{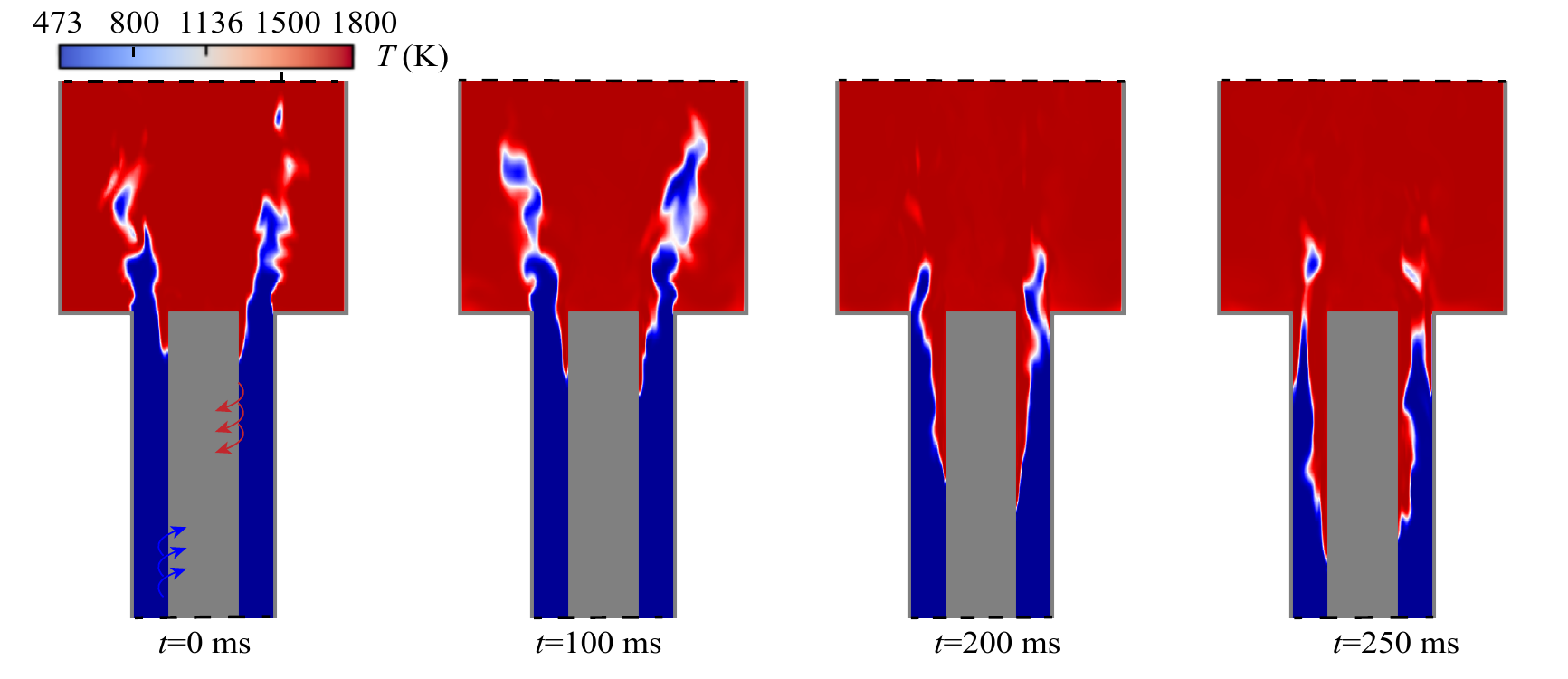}
	\caption{
        Instantaneous contours of $\widetilde{c}$ in LES-FSD at different times in case E1 with $X_\mathrm{H_2}=80\%$, $ \phi=0.558$, $p=2.5$ bar, $T_{ub}=473$ K, and $U_x=15$ m/s.
    }
	\label{fig:flameshape}
\end{figure}

First, we present the transient BLF process with the rotating flame tongue from the LES-FSD of case E1 with $X_\mathrm{H_2}=80\%$, $p=2.5$ bar, and $\phi=0.558$.
Figure~\ref{fig:flameshape} depicts instantaneous contours of $\tilde{c}$ on the $x$--$r$ plane at $t=0$, 100, 200, and 250 ms, where $t=0$ ms marks the beginning of BLF.
Directions of flame propagation and swirling flow are marked by red and blue arrows, respectively.
The flame propagates upstream asymmetrically along the central bluff body, as a large-scale flame tongue rotating around the central bluff body.
The leading point of the flame, or the flame base, is quantified by the lowest position of the isosurface of $\widetilde{c}=0.68$.
This isosurface has the maximum heat release rate in the corresponding laminar flame.

\begin{figure}
	\centering
	\includegraphics[width=0.5\textwidth]{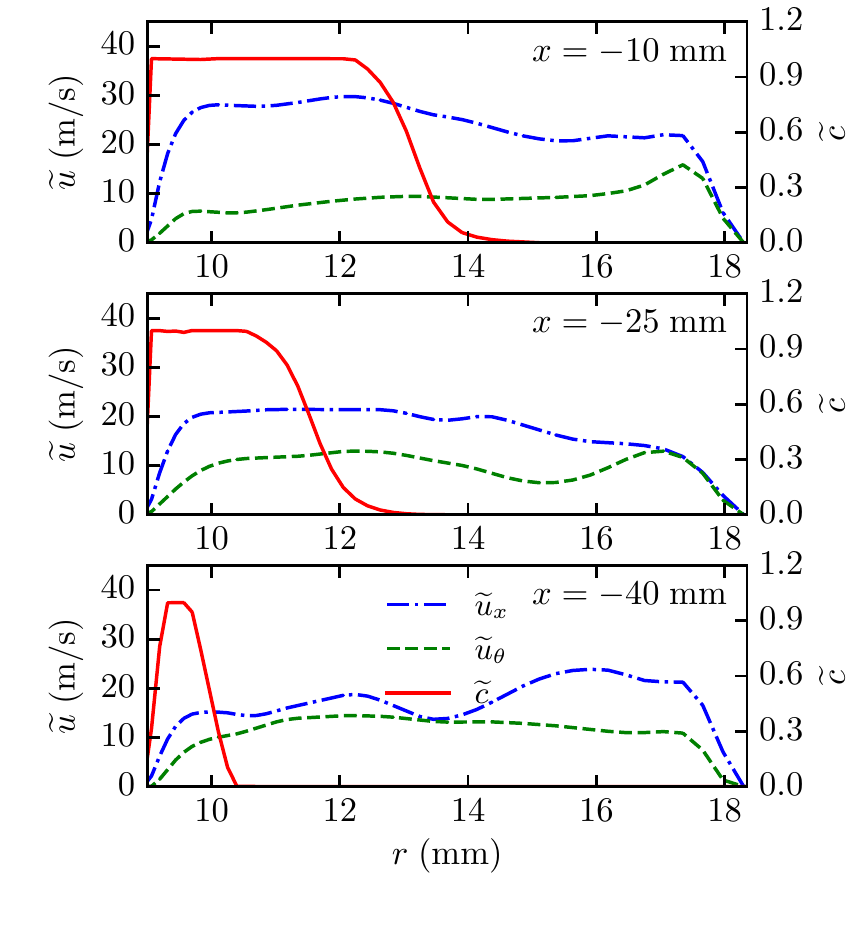}
	\caption{
        Radial profiles of $\widetilde{u}_x$, $\widetilde{u}_\theta$, and $\widetilde{c}$ at $t=200$ ms and $x=-10$, $-25$, and $-40$ mm in case E1 with $X_\mathrm{H_2}=80\%$, $ \phi=0.558$, $p=2.5$ bar, $T_{ub}=473$ K, and $U_{x}=15$ m/s.
        }
	\label{fig:inst_flame}
\end{figure}

Figure~\ref{fig:inst_flame} plots radial profiles of axial velocity $\widetilde{u}_x$, azimuthal velocity $\widetilde{u}_\theta$, and $\tilde{c}$ at $t=200$ ms and $x=-10$, $-25$, and $-40$ mm, where the $\theta$-coordinate is adjusted by rotating the flame base onto the $x$--$r$ plane at $\theta=0$.
The overall azimuthal velocity slightly decays downstream due to the friction drag.
The axial motion is accelerated as passing through the flame within $r<14~\mathrm{mm}$ at $x=-40$ mm.
The leading edge of the flame tongue, represented by the peak of the $\tilde{c}$ profile, stays in the boundary layer of the central bluff body.

\begin{figure}
	\centering
	\includegraphics[width=\textwidth]{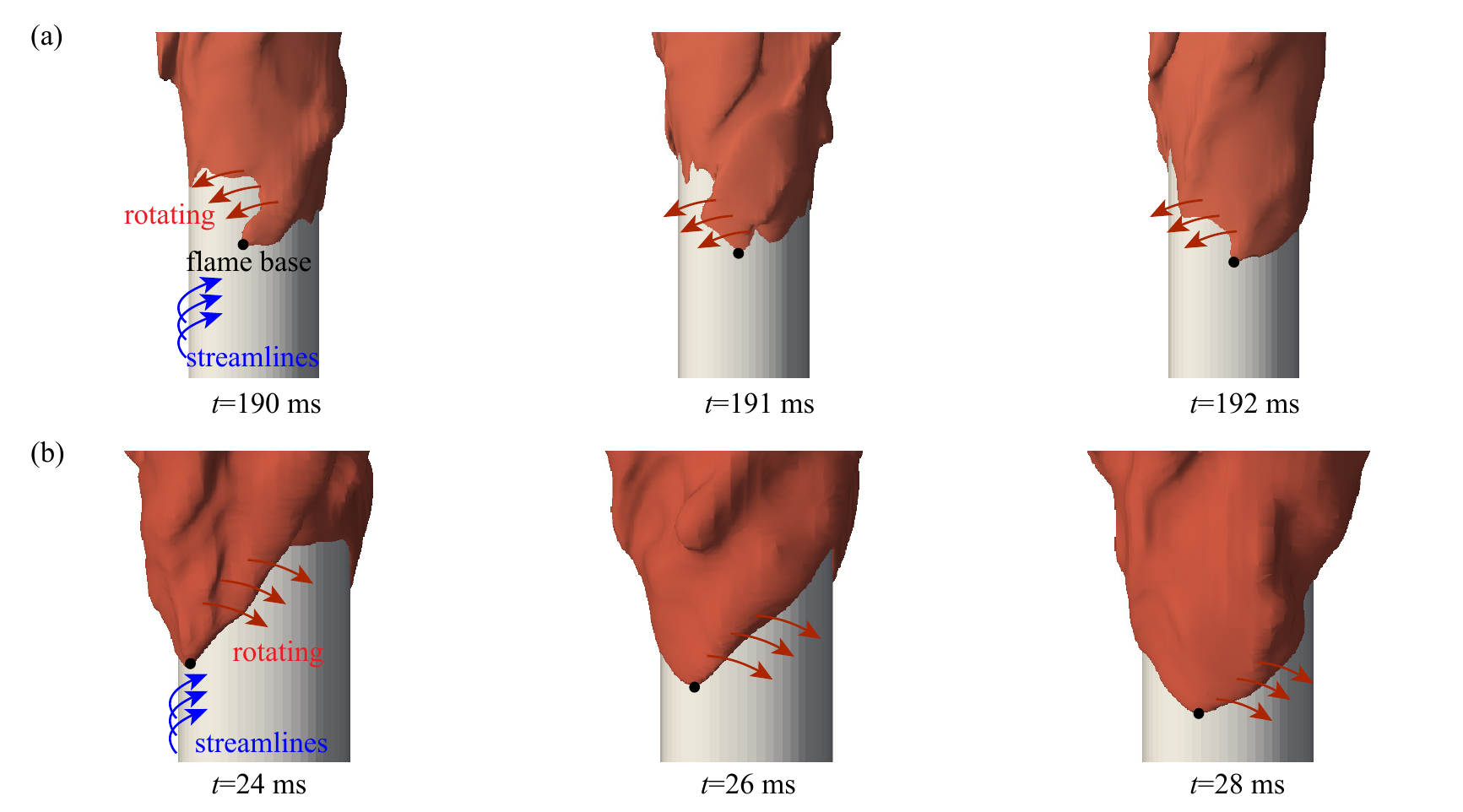}
	\caption{
        Propagation of the flame tongue during the BLF of (a) $\mathrm{CH_4}$/$\mathrm{H_2}$/air flames with $X_\mathrm{H_2}=80\%$, $\phi=0.558$, $p=2.5$ bar, $T_{ub}=473$ K, $S=0.7$, and $U_x=15$ m/s and (b) $\mathrm{CH_4}$/air flames with $\phi=1.0$, $p=1.0$ bar, $T_{ub}=293$ K, $S=0.9$, and $U_x=5$ m/s.
    }
	\label{fig:flamepathways}
\end{figure}

We compare the evolutions of the isosurface of $\tilde{c}=0.68$ during the BLF in case E1 in Fig.~\ref{fig:flamepathways}a and in another CH$_4$/air flame with $\phi=1.0$ and $p=1.0$ bar in Fig.~\ref{fig:flamepathways}b.
The latter case corresponding to the experiment~\cite{Ebi2016} in a similar combustor is detailed in \ref{sec:atmospheric_pressure}.
Directions of flame propagation and swirling flow are marked by red and blue arrows, respectively.
The BLF propagates as a large-scale flame tongue. 
In Fig.~\ref{fig:flamepathways}a for case E1, the flame tongue rotates against the swirl flow, and we observe the same flame propagation pathway for all of the simulated hydrogen-enriched flame cases at $p=2.5$ bar.
By contrast, the flame tongue tends to rotate along with the bulk flow in the swirling flames at $p=1.0$ bar.
We refer the former and latter propagation modes to ``upwind'' and ``crosswind'', respectively.
The flame-tongue structure and the two propagation modes in the present LES-FSD agree well with experimental observations~\cite{Ebi2016,Ebi2018,Ebi2021}.

\begin{figure}
	\centering
	\includegraphics[width=0.5\textwidth]{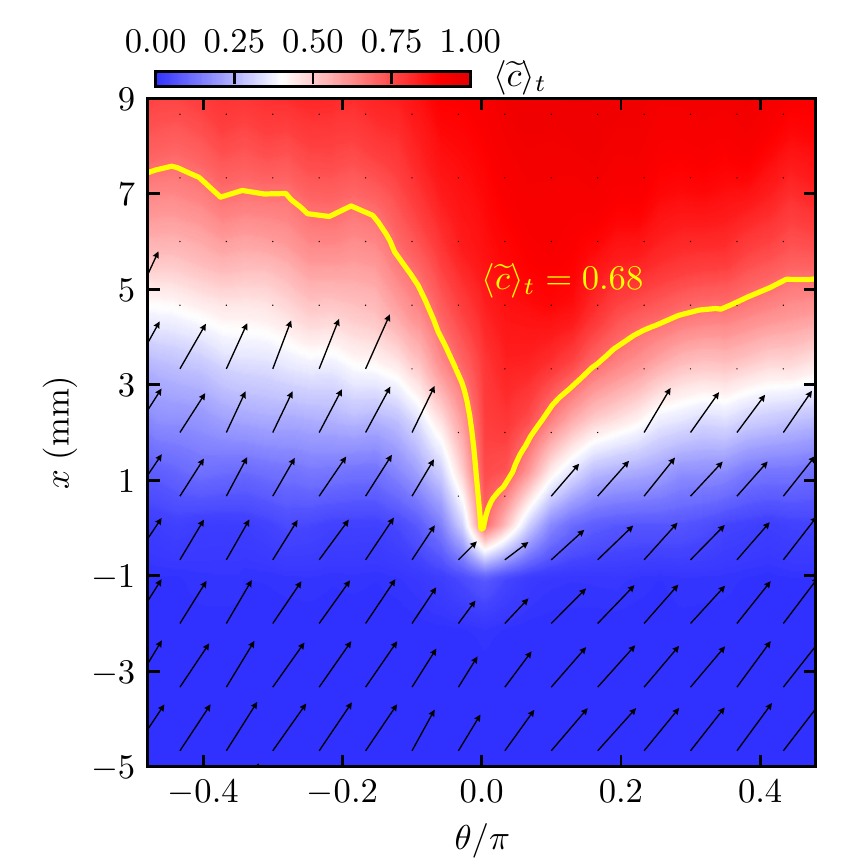}
	\caption{
        Time-averaged flame front and velocity vectors over a period from $t=100$ ms to 200 ms in case E1. The arrows denote $\langle \widetilde{u}_s \rangle_t$.
	}
	\label{fig:vector}
\end{figure}

To obtain the overall flame geometry, we take the time average of the flow field in case E1 from $t=100$ to 200 ms.
During this period, the BLF is considered in a steady state. It propagates with a constant speed in axial and azimuthal directions.
To average the transient flashback, the flame base at each time are translated in the axial and azimuthal axes to $x=0$ and $\theta=0$.
Figure~\ref{fig:vector} depicts the contour of $\langle \widetilde{c} \rangle_t$ and vectors of $\langle \widetilde{u}_s \rangle_t$ on the unwrapped $x$--$\theta$ plane, where $\langle\cdot\rangle_t$ denotes the time average and $\widetilde{u}_s$ is the velocity projected onto the $x$--$\theta$ plane.
The contour line of $\langle \widetilde{c} \rangle_t = 0.68$ illustrates a convex flame front, i.e., the flame tongue, with one side across and the other one aligned with the swirling flow direction.
The two sides are referred to as the upwind and crosswind sides, respectively.
Thus, the side propagating upstream determines the propagation mode of BLF.

\begin{figure}
	\centering
	\includegraphics[width=0.5\textwidth]{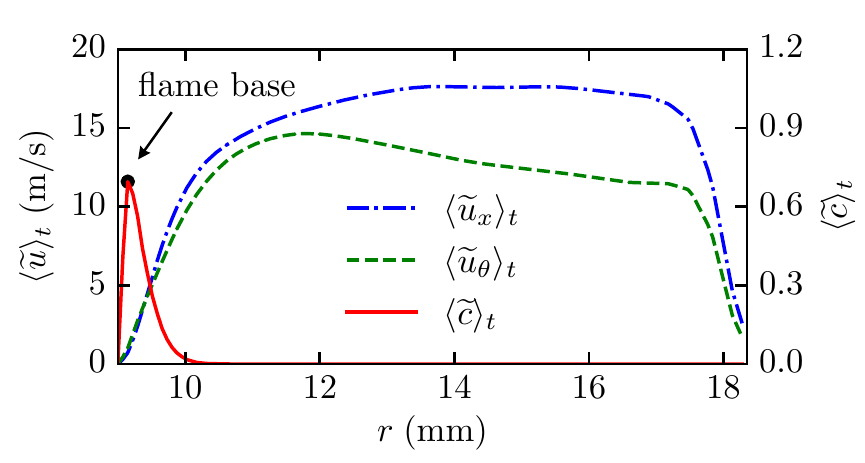}
	\caption{
		Radial profiles of $\langle \widetilde{u}_x \rangle_t$, $\langle \widetilde{u}_\theta \rangle_t$, and $\langle \widetilde{c} \rangle_t$ crossing the flame base in case E1. 
	}
	\label{fig:statistic}
\end{figure}

In Fig.~\ref{fig:statistic}, radial profiles of $\langle \widetilde{u}_x \rangle_t$, $\langle \widetilde{u}_\theta \rangle_t$, and $\langle \widetilde{c} \rangle_t$ of the time-averaged flame tongue demonstrate that the flame base is in the boundary layer of the central bluff body.
The decrease of $\langle \widetilde{c} \rangle_t$ near wall is caused by the flame quenching with the heat loss through the boundary.
This confirms again that the flashback in the present simulations is driven by the BLF.

\begin{figure}
	\centering
	\includegraphics[width=0.5\textwidth]{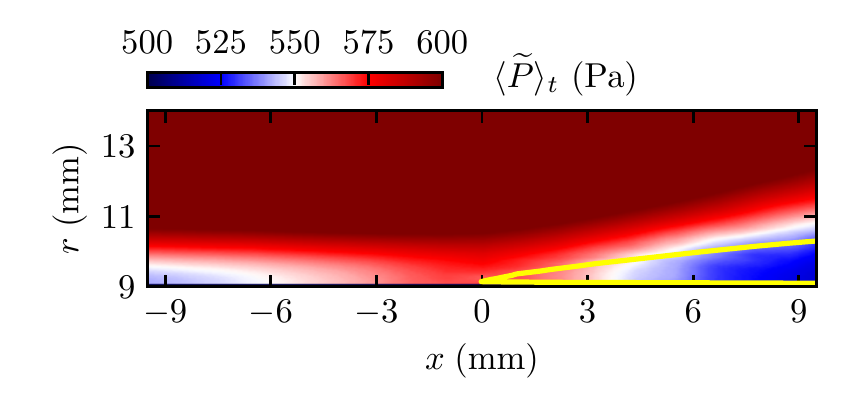}
	\caption{
        Time-averaged pressure from 100 to 200 ms on the $x$-$r$ plane crossing the flame base in case E1.
	}
	\label{fig:pressure}
\end{figure}

The propagation pathway and radial profiles indicate that the BLF of swirling flames at high pressures is similar to the BLF in non-swirling flames. 
To further support the observation, Fig.~\ref{fig:pressure} plots the time-averaged pressure on the $x$--$r$ plane crossing the flame base in case E1.
The contour line of $\langle \widetilde{c} \rangle_t = 0.68$ illustrates the shape of the flame front. 
It is observed that the high pressure zone is near the leading edge of the flame front, and the pressure decreases downstream the flame. 
This is similar to the DNS results on BLF in non-swirling flows~\cite{Gruber2012}, where the elevated pressure zone is restricted to the vicinity of small-scale bulges.

\subsection{BLF limit of swirling flames}\label{sec:LES_FL}

\begin{figure}
	\centering
	\includegraphics[width=0.5\textwidth]{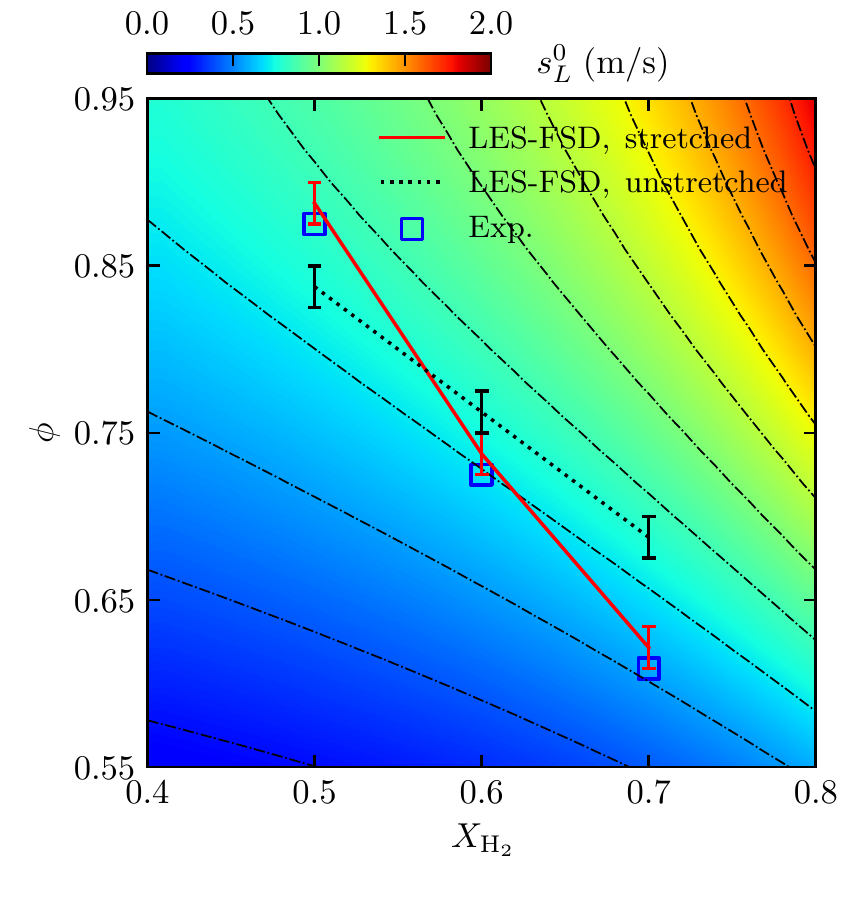}
	\caption{
        Comparisons of the BLF limits obtained in LES-FSD (lines) and experiments~\cite{Ebi2021} (symbols) in cases A1, A2, and A3 with $U_x=15$ m/s and different $X_{\mathrm{H}_2}$.
        The red solid and black dotted lines represent LES-FSD results using models of $\langle s_d\rangle_A$ with and without flame stretch effects, respectively.
        The upper and lower bounds of the error bar with the width $\Delta\phi=0.025$ denotes the values of $\phi$ in flashback and flame stabilization, respectively.
        }
	\label{fig:flamelimit}
\end{figure}

We examine the BLF limit of various swirling flames.
Following the procedure described in Section~\ref{sec:config}, the critical equivalence ratio $\phi_{cr}$ is determined by two simulations of flame stabilization and flashback with a small difference of $\phi$.
Figure~\ref{fig:flamelimit} compares $\phi_{cr}$ obtained from the LES-FSD with the $\langle s_d\rangle_A$ model in Eq.~\eqref{eq:sdm2} considering flame stretch effects (red solid line) and the experiment (symbols)~\cite{Ebi2021} for cases A1, A2, and A3 with the same inlet bulk velocity $U_x=15$ m/s.
The upper and lower bounds of the error bar are $\phi$ in flashback and flame stabilization cases, respectively.
The width of the error bar denotes the equivalence ratio increment $\Delta\phi=0.025$. 
The contour is color-coded by $s_L^0$ in terms of $X_{\mathrm{H}_2}$ and $\phi$.

The results of the LES-FSD with the $\langle s_d\rangle_A$ model in Eq.~\eqref{eq:sdm2} and experiments agree well at different $X_{\mathrm{H}_2}$.
As $X_{\mathrm{H}_2}$ grows from 50\% to 70\%, $\phi_{cr}$ decreases from 0.875 to 0.609, and the corresponding $s_L^0$ decreases from 0.815 m/s to 0.591 m/s.
The misalignment between the decaying profile of $\phi_{cr}$ and the contour line of $s_L^0$ indicates that the turbulence and flame stretch play important roles in hydrogen-enriched flames.
With the present $\langle s_d \rangle_A$ model incorporating the flame stretch effect in Eq.~\eqref{eq:sdm2}, the LES-FSD well estimates the flashback limits.

\begin{figure}
    \centering
	\includegraphics[width=0.45\textwidth]{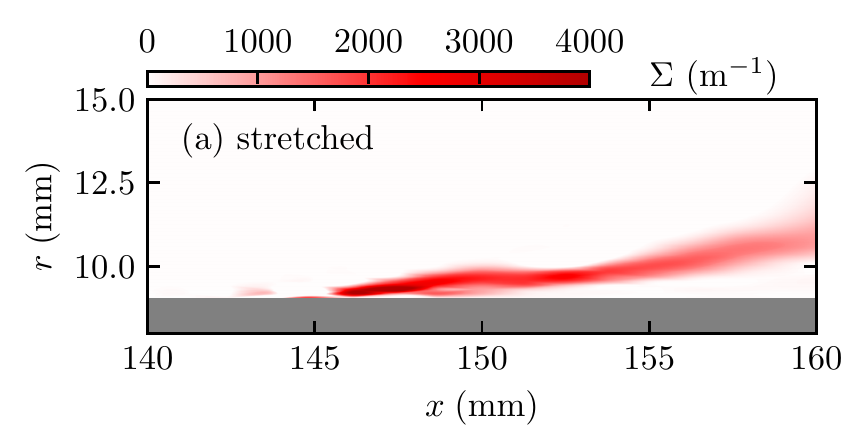}
	\includegraphics[width=0.45\textwidth]{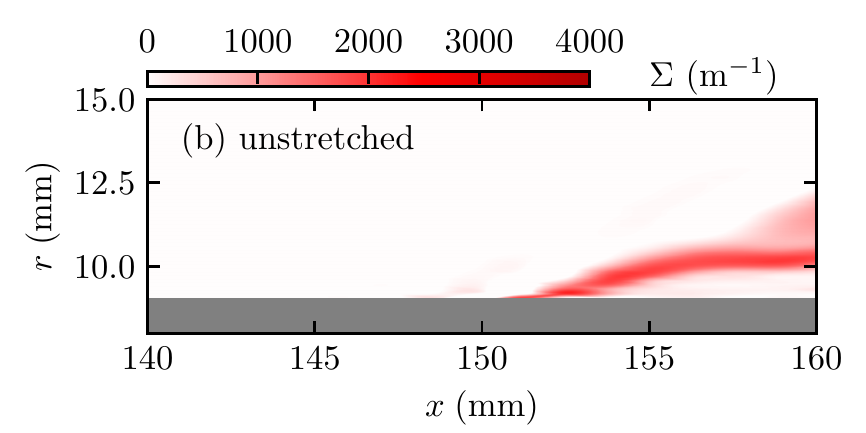}
	\caption{Instantaneous contours of $\Sigma$ of case A3 in LES-FSD (a) with and (b) without stretch effects modeled via $I_0$.}
	\label{fig:FSD}
\end{figure}

Moreover, the flashback limit obtained from the LES-FSD with the $\langle s_d\rangle_A$ model in Eq.~\eqref{eq:sdm1} based on unstretched flames (black dotted line) shows a notable discrepancy from the experiment result in Fig.~\ref{fig:flamelimit}.
Since the inlet flow velocities are the same, we assume that the turbulent burning velocity $s_T$ at the BLF limit is constant. 
Thus, LES-FSD with the stretch effects gives larger $s_T/s_L^0$ at lower equivalence ratios. 
Figure~\ref{fig:FSD} compares $\Sigma$ obtained from LES-FSD with and without the stretch effects for case A3.
It shows that $\Sigma$ is larger near the flame front when the stretch effects are modeled, indicating that the flame wrinkling is enhanced due to the thermo-diffusive effects.

\begin{figure}
	\centering
    \includegraphics[width=\textwidth]{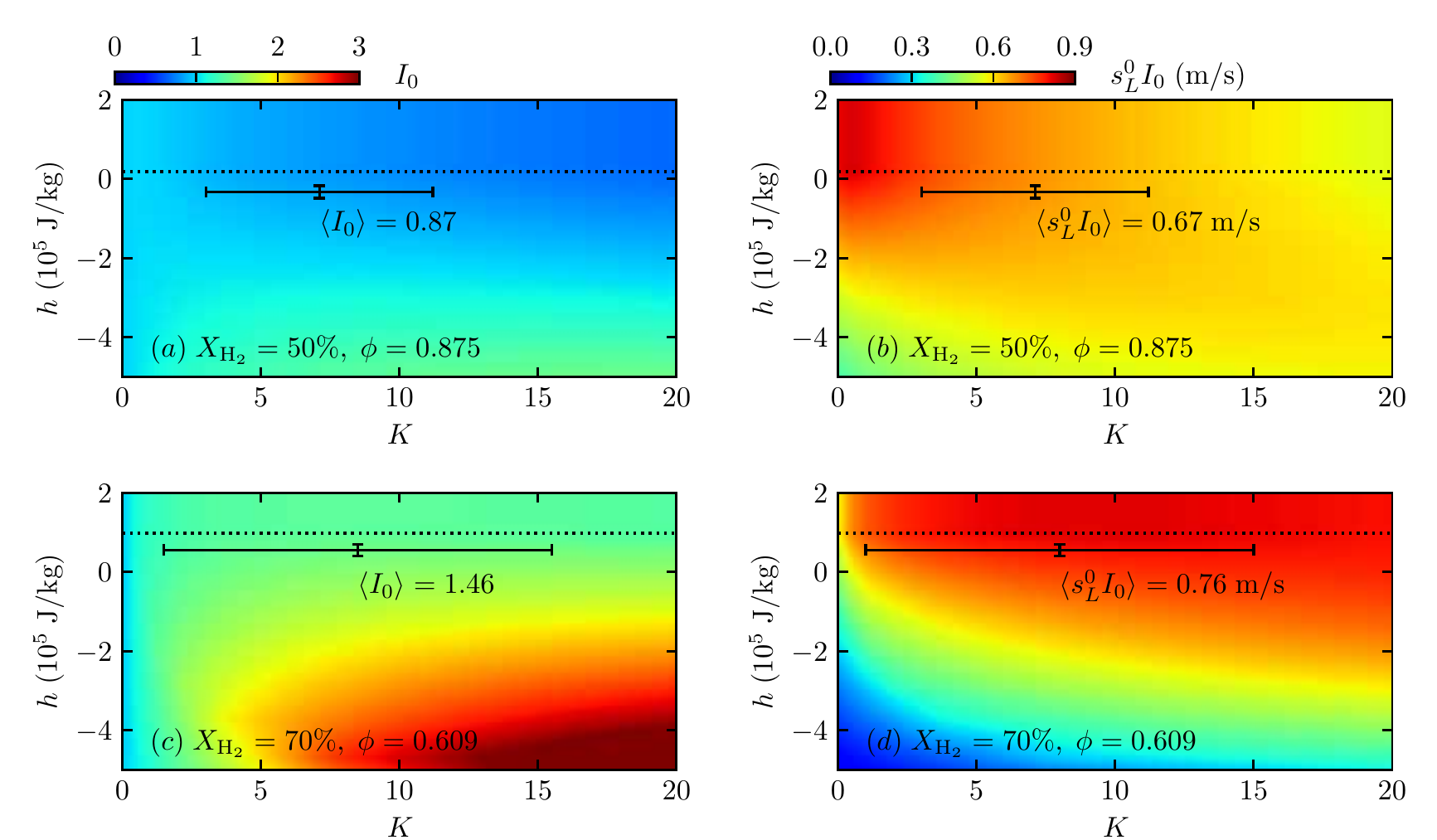}
	\caption{
        Contours of $I_0$ and $s_L^0 I_0$ in terms of $K$ and $h$ in cases A1 (upper row) and A3 (lower row) at the BLF limit $\phi=\phi_{cr}$. The point with error bars presents averaged values with one standard deviations of $K$ and $h$ on the leading edge of the flame front.
	}
	\label{fig:I0_HL}
\end{figure}

The enhanced flame stretch effects due to the hydrogen-enrichment and low equivalence ratio lead to the misalignment between the decaying trends of the BLF limit and $s_L^0$ in Fig.~\ref{fig:flamelimit}.
Figure~\ref{fig:I0_HL} plots $I_0$ and $s_L^0 I_0$
in terms of $K$ and $h$ in cases A1 with $X_{\mathrm{H}_2}=50\;\%$ and A3 with $X_{\mathrm{H}_2}=70\;\%$ at the BLF limit $\phi=\phi_{cr}$.
The point with error bars presents averaged values with one standard deviations of $K$ and $h$ on the leading edge of the flame front.
For the adiabatic condition (marked by the dotted line), $I_0$ in case A1 is close to unity and decreases slightly with $K$, and $I_0$ in case A3 rises with $K$ up to values around two, because case A3 with a larger $X_{\mathrm{H}_2}$ and a lower $\phi_{cr}$ is associated with stronger thermo-diffusive effects.
Considering the heat loss, both cases show that $I_0$ increases with the decrease of $h$, although $s_L^0$ decreases with $h$.
%
The thermo-diffusive effects lead to larger $s_T/s_L^0$ for case A3. 
This agrees with Fig.~\ref{fig:flamelimit} that the BLF happens at lower $s_L^0$ for the higher hydrogen-enrichment level with stronger thermo-diffusive effects.

In summary, the large $s_L^0$ of hydrogen leads to the decrease of $\phi_{cr}$ with increasing $X_{\mathrm{H}_2}$ to stabilize the flame at the fixed $U_x$.
At the same time, the hydrogen-enrichment and low equivalence ratio bring strong thermo-diffusive effects, accelerating turbulent flame propagation.
Consequently, a lower $\phi_{cr}$ is required to stabilize the hydrogen-enriched flames, which is represented as the misalignment between the decaying trends of the BLF limit and $s_L^0$ in Fig.~\ref{fig:flamelimit}.
The LES-FSD with the $\langle s_d\rangle_A$ model in Eq.~\eqref{eq:sdm2} captures this phenomenon through the modeling of the flame stretch effects.

\section{Modeling of the BLF limit}\label{sec:FLmodel}

\subsection{BLF modes}\label{sec:pathways}

\begin{figure}
	\centering
	\includegraphics[width=0.5\textwidth]{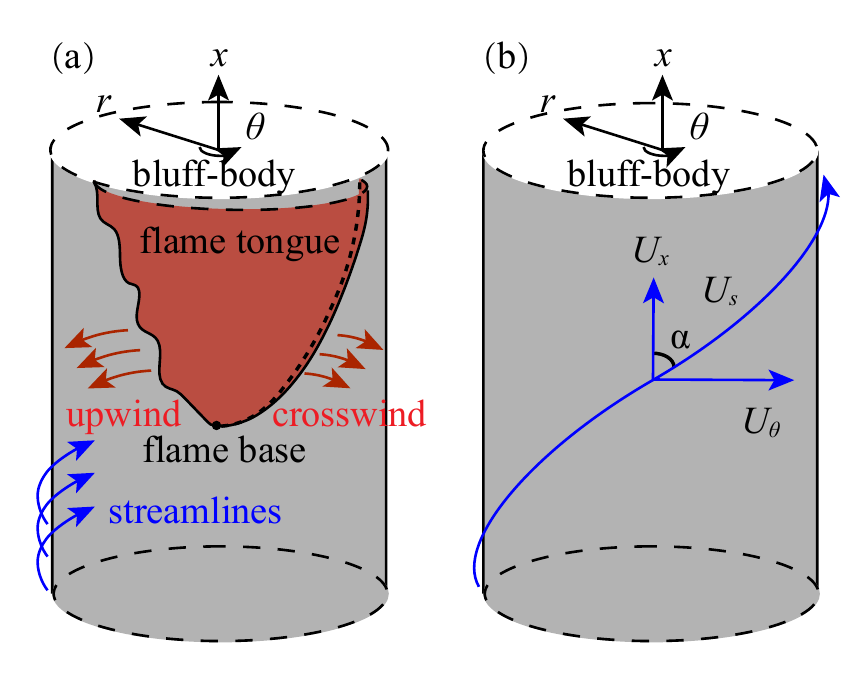}
	\caption{
        Schematics of (a) propagation modes of the flame tongue in the mixing tube and (b) the velocity decomposition along the direction of the bulk flow.
        }
	\label{fig:flametongue}
\end{figure}

In order to illustrate the different modes of flame propagation during the BLF, the propagating flame tongue in the mixing tube is sketched in Fig.~\ref{fig:flametongue}a, where the flame and central bluff body are represented as red and gray surfaces, respectively.
The flame base divides the flame tongue into the upwind and crosswind sides.
The propagation mode is determined by the dominant side propagating upstream.
From the overall shape of the flame tongue in Fig.~\ref{fig:vector}, we approximate that the leading edge of the flame front on the upwind side is normal to the bulk flow direction and is parallel on the crosswind side.
Then, we estimate the angle (marked in Fig.~\ref{fig:flametongue}b)
\begin{equation}\label{eq:angle}
    \alpha = \arctan\left[ 1.5 S (R^3_2 - R_1^2 R_2)/(R_2^3 - R_1^3) \right]
\end{equation}
between the bulk flow and the axial direction based on the definition of the swirl number~\cite{Vignat2022}, where $R_1$ and $R_2$ are the inner and outer radii of the mixing tube, respectively.
A validation on the approximation of $\alpha$ in Eq.~\eqref{eq:angle} is given in~\ref{sec:atmospheric_pressure}.

\begin{figure}
	\centering
	\includegraphics[width=\textwidth]{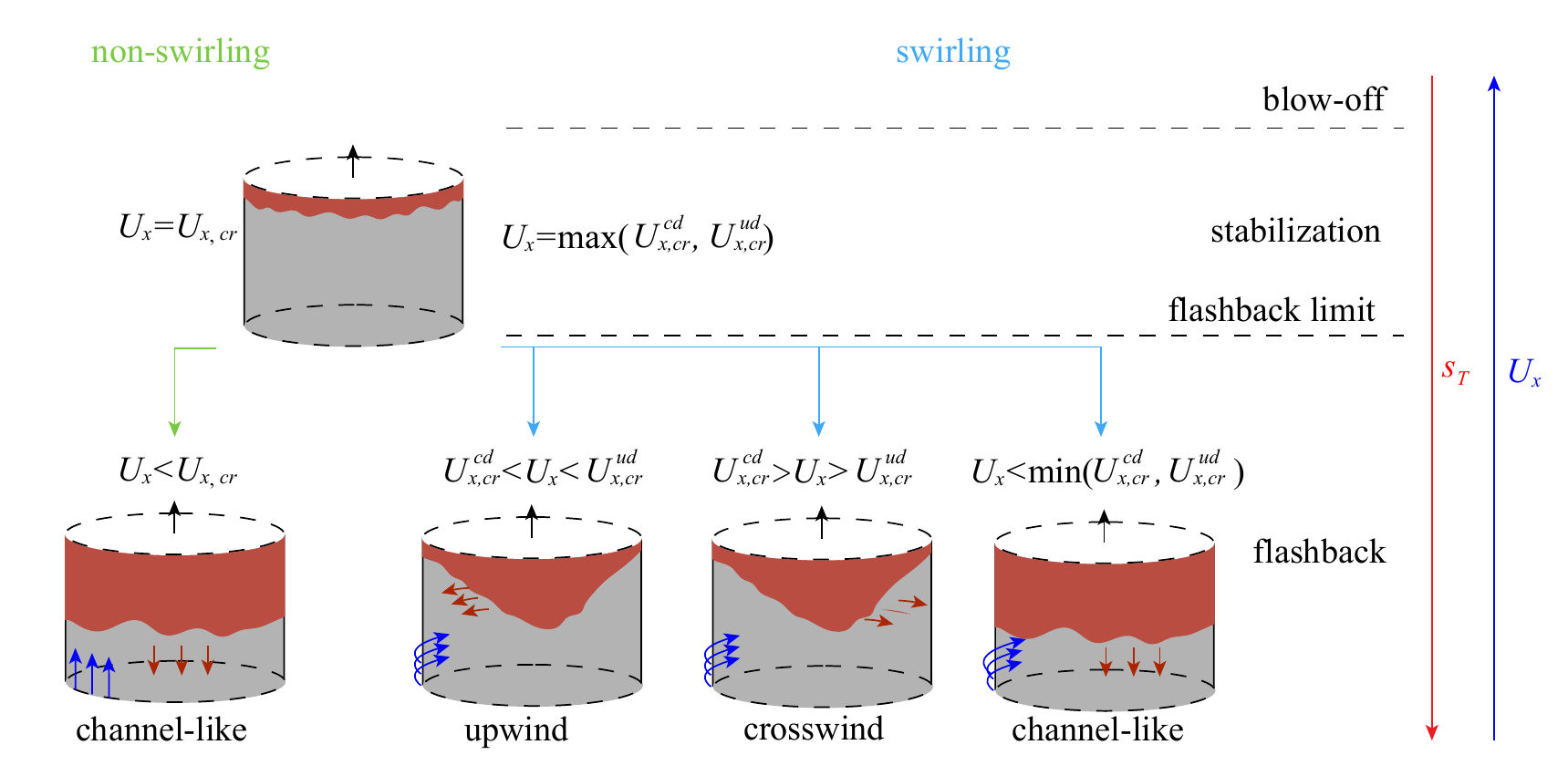}
	\caption{Schematics on the modes of flame stabilization and BLF in the mixing tube.}
	\label{fig:flashbackmode}
\end{figure}

For each side of the flame tongue, a critical axial bulk velocity $U_{x,cr}$ is calculated at the BLF limit.
The BLF happens for $U_x < U_{x,cr}$; otherwise, the flame is stabilized or blow-off.
The different modes of BLF in the mixing tube are sketched in Fig.~\ref{fig:flashbackmode}.
For the non-swirling bulk flow, the bugles of BLF propagate along the central bluff body~\cite{Hoferichter2017}, similar to the BLF in channels~\cite{Gruber2012}.
For the swirling flames, the propagating pathway of BLF depends on the values of $U_{x,cr}^{ud}$, $U_{x,cr}^{cd}$, and $U_x$, 
where the superscripts $ud$ and $cd$ denote $U_{x,cr}$ of the upwind and crosswind sides of the flame tongue, respectively.
For $U_{x,cr}^{cd} < U_x < U_{x,cr}^{ud}$, the BLF occurs on the upwind side only, and the flame tongue rotates against the bulk flow.
For $U_{x,cr}^{cd} > U_x > U_{x,cr}^{ud}$, the BLF occurs on the crosswind side only, and the flame tongue swirls along the direction of the bulk flow.
If $U_{x}$ is less than both $U_{x,cr}^{cd}$ and $U_{x,cr}^{ud}$, both the upwind and crosswind sides propagate upstream, and the BLF is similar to the channel-like mode in the non-swirling flame, which was observed in the LES of the swirling flame with an adiabatic central bluff body~\cite{Xia2022}.

Based on the projection of the velocity with the angle $\alpha$, we have $U_{x,cr}^{cd}/U_{x,cr}^{ud}=\tan\alpha$.
In the present study, all cases at 2.5 bar have $\tan\alpha=0.90$, with $U_{x,cr}^{cd} < U_{x,cr}^{ud}$ for the upwind flashback.
Meanwhile, the cases at 1 bar have $\tan\alpha=1.16$, with $U_{x,cr}^{cd} > U_{x,cr}^{ud}$ for the crosswind flashback.
The two different flashback modes are observed in the LES-FSD simulations and experiments~\cite{Ebi2016,Ebi2021}. 
The hydrogen-enriched cases at $p=2.5$ bar show a flame tongue rotating against the swirl flow, whereas it rotates along with the bulk flow in the CH$_4$/air flame at $p=1.0$ bar. 
Two examples of each flashback are presented in Fig.~\ref{fig:flamepathways}.

\subsection{Prediction of the BLF limit}\label{sec:Predictmodel}

It is useful to predict the BLF limit of swirling flames with different hydrogen additions using a simple model in practical applications.
Regarding the flame front characteristics as discussed in Sec.~\ref{sec:flameshape}, we model the BLF limit for the upwind flashback mode as in non-swirling flows.
Here we extend the model of Hoferichter \etal~\cite{Hoferichter2017} by incorporating the BLF mode analysis in Section~\ref{sec:pathways} and the recently developed model of the turbulent burning velocity $s_T$~\cite{Lu2022}.

Based on the boundary-layer separation, Hoferichter \etal~\cite{Hoferichter2017} estimated the BLF limit in channels via a momentum balance of the incoming flow and the pressure rise induced by turbulent flame propagation as
\begin{equation}\label{eq:original_model}
	U_{x, cr} + 2.4 u_{\tau, cr}
	=
	s_T
	\sqrt{ 4.721 \left( \frac{\rho_{ub}}{\rho_b} -1 \right)},
\end{equation}
where $u_{\tau, cr} = 0.18 \left( {U_{x,cr} H}/{\nu} \right)^{0.88} \nu/H$ is the shear stress velocity~\cite{Pope2000}, $\nu$ is the viscosity, and $H$ is the channel height.
This model employs a power law of $s_T$ with a linear dependence on the flame stretch.
Although the model was validated for hydrogen-air flames in confined channels, it needs to be improved for hydrogen-enriched swirling flames~\cite{Ebi2021}.

From the propagating direction of the flame tongue in Fig.~\ref{fig:flametongue}, we project $s_T$ in the axial direction onto the upwind side as
\begin{equation}\label{eq:model}
	U_{x,cr}^{ud} + 2.4 u_{\tau, cr}
	= s_T \cos \alpha \sqrt{ 4.721 \left( \frac{\rho_{ub}}{\rho_b} -1 \right)}.
\end{equation}
Here, $\alpha$ is calculated by Eq.~\eqref{eq:angle} which involves the effects of the swirl number and geometry of the mixing tube.
Note that the flame front of the upwind side may not be exactly perpendicular to the averaged bulk flow direction.
Consequently, the angle between the BLF propagation and axial directions can differ from $\alpha$, as observed in Fig.~\ref{fig:vector}.
To quantify the uncertainty introduced by this assumption, we tested the model with $\alpha$ obtained in LES-FSD simulations. 
The small mean discrepancy 4.16\% between $\alpha$ calculated by Eq.~\ref{eq:angle} and measured in experiments~\cite{Ebi2016} is acceptable. 

On the crosswind flashback mode, experiments~\cite{Karimi2015,Ebi2016} observed the raise of pressure in downstream combustion zone.
It was explained by the effects of the centripetal force of the swirling flows~\cite{Karimi2015}. 
A recent DNS on planar channel flow~\cite{Bailey2021} showed that the wall-normal pressure gradients can induce a streamwise pressure difference, thus increasing the boundary-layer flashback speed. 
Therefore, the centripetal force affects the BLF limit for the crosswind mode, i.e., $U_{x,cr}^{cd}$. 
Further investigations are needed to develop a more general BLF model for the crosswind flashback mode.

The model of $s_T$ of Lu and Yang~\cite{Lu2022} gives
\begin{equation}
	\dfrac{s_T}{s_L^0} =
	I_0
	\exp\!
    \left\{\!
	\left[\! T_\infty^*
        \left(\mathcal{A}\!+\!\mathcal{B}s^0_{L0} I^2_0\right)
        \!+\!\frac{1}{2}\ln\left(\frac{l_T}{\delta^0_{L}}\right)
	\!\right]
	\left[
	1\!-\!\exp\left(
    -\dfrac
	{	
        \mathcal{C}
		Re ^{-\frac{1}{4}}
		\left( l_T / \delta^0_L \right)^{\frac{1}{2}}
	}
	{T_\infty^* \left( \mathcal{A} \!+\! \mathcal{B} s^0_{L0} I^2_0\right)I_0}
	\frac{u^\prime}	{s^0_{L}}
	\right)
	\right]
	\!\right\},
	\label{eq:stmodel}
\end{equation}
where $\mathcal{A}=0.317$, $\mathcal{B}=0.033$, and $T_\infty^*=5.5$ are universal constants determined by Lagrangian statistics in non-reacting homogeneous isotropic turbulence~\cite{You2020},
the dimensionless laminar flame speed $s^0_{L0} = s^0_{L} /s_{L,ref}$ is normalized by a reference value $s_{L,ref} = 1$ m/s,
$\mathcal{C} = (1-\rho_b/\rho_{ub})I_0(K=1)/Le$ is a fuel-dependent coefficient,
$Le$ is the Lewis number, $l_T$ is the turbulent integral length,
$u'$ is the turbulence intensity,
and $Re = u' l_T / \nu $ is the turbulence Reynolds number.
The validations against a number of DNS/experimental datasets demonstrated that the model in Eq.~\eqref{eq:stmodel} works well for a wide range of conditions, including the hydrogen and hydrogen-enriched flames at high pressures~\cite{Lu2020,Lu2022}.
In the present application, we set $l_T$ to be 7\% of the hydraulic diameter and $u'=2.6 u_\tau$~\cite{Hoferichter2017}.
Comparing with the $s_T$ model for adiabatic flames,
we account for the heat loss through $s_L^0$ in Eq.~\eqref{eq:stmodel}.
To look up $s_L^0(\widetilde{h}, \widetilde{Z})$, the loss of the sensible enthalpy is estimated to be 10\%, based on the enthalpy statistics at the flame front obtained via the LES-FSD of case E1.

Predicting the flashback limit can help the design of premixed swirling burner.
Here, we estimate $U_{x, cr}^{ud}$ for given $X_{\mathrm{H}_2}$ and $\phi$ from Eqs.~\eqref{eq:model} and \eqref{eq:stmodel}.
Alternatively, $\phi_{cr}$ can be obtained for given $X_{\mathrm{H}_2}$ and $U_x$.
The model of the BLF limit is assessed by the experimental results of swirling flames in Ebi \etal~\cite{Ebi2021}.
As listed in Table~\ref{tab:exp_cases}, the experiments with $p=2.5$ bar and $T_{ub}=473$ K cover a range of operating conditions of $X_{\mathrm{H}_2}$, $\phi$, and $U_x$.

\begin{table}
	\caption{Comparisons of experimental results and model predictions of $U_{x,cr}^{ud}$ in CH$_4$/H$_2$/air swirling flames.}
    \centering
	\begin{tabular}{ccccc}
		\hline
		case    &$X_\mathrm{H_2}\;(\%)$ &$\phi$	& $U_{x,cr}^{ud}\;(\mathrm{m/s})$, exp. & $U_{x,cr}^{ud}\;(\mathrm{m/s})$, model \\
		\hline
		A1		&50	&0.875	&15 	&14.50	\\
		A2		&60	&0.725  &15 	&16.94	\\	
		A3		&70	&0.609  &15 	&15.91	\\
        A4		&100&0.353  &15 	&12.85	\\
		B1		&60	&0.855  &20 	&19.85  \\
		B2		&70	&0.687	&20 	&22.02	\\
		B3		&80	&0.558  &20 	&19.45	\\
        B4		&100&0.372  &20 	&20.35	\\
		C1		&70	&0.795  &25 	&27.23	\\
		C2		&80	&0.608  &25 	&25.14	\\
		C3		&85	&0.535  &25 	&22.32	\\
		D1		&80	&0.709  &30 	&35.89	\\
		D2		&85	&0.628	&30 	&35.52	\\	
        D3		&100&0.421  &30 	&30.77	\\	
		\hline
	\end{tabular}
	\label{tab:exp_cases}
\end{table}

\begin{figure}
	\centering
	\includegraphics[width=0.45\textwidth]{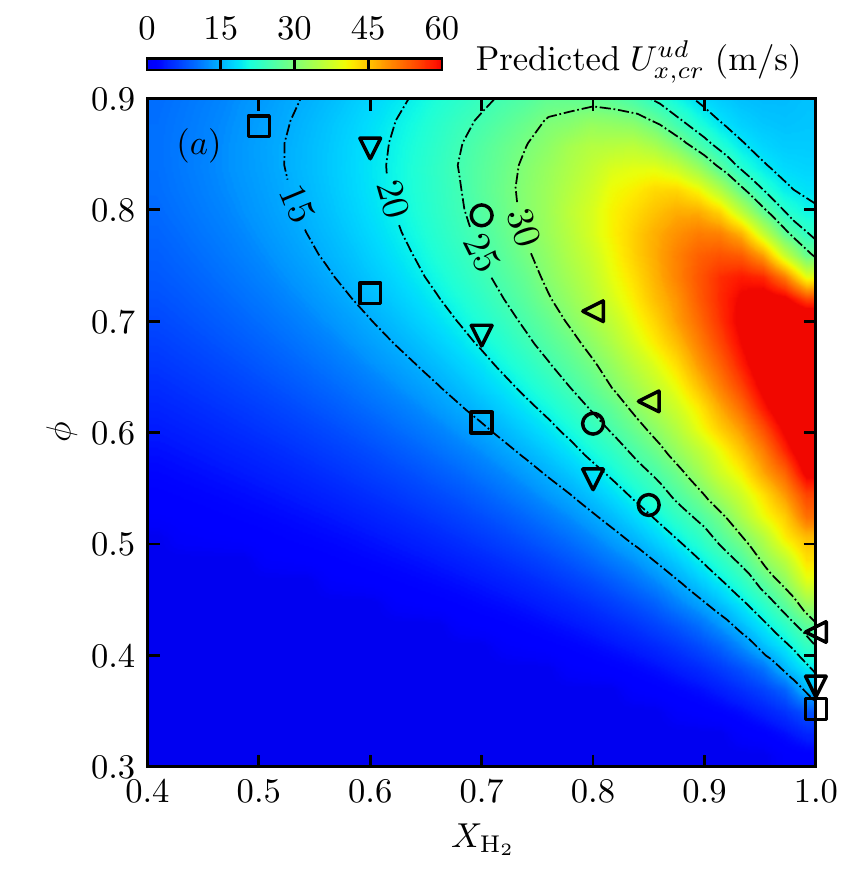}
	\includegraphics[width=0.45\textwidth]{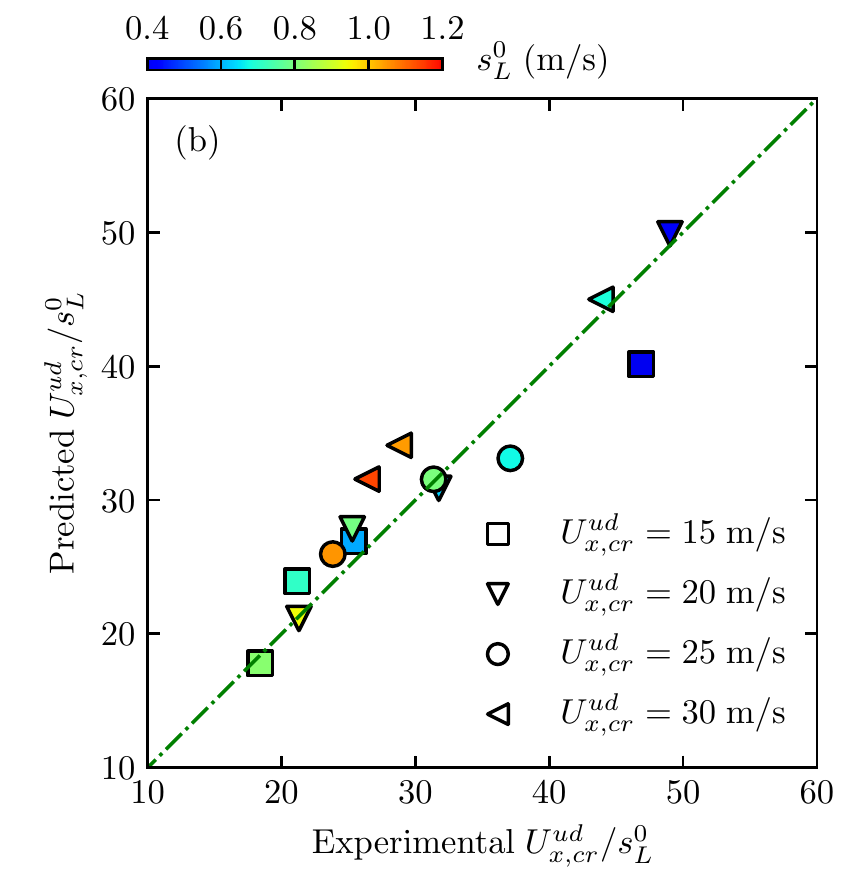}
	\caption{
        Comparisons on the BLF limit between the experiment~\cite{Ebi2021} and the simple model in Eqs.~\eqref{eq:model} and \eqref{eq:stmodel}.
        (a) Contour of $U_{x,cr}^{ud}$ in terms of $X_{\mathrm{H}_2}$ and $\phi$ is calculated by the model, and symbols denote experimental results with values of $U_x$ provided in legends in (b).
        (b) Comparison between experimental and modeling results of $U_{x,cr}^{ud}$.
    }
	\label{fig:model_assess}
\end{figure}

Figure~\ref{fig:model_assess}a compares the model predictions and experimental results in terms of $X_\mathrm{H_2}$ and $\phi$.
The contour of $U_{x,cr}^{ud}$ is calculated using the model in Eqs.~\eqref{eq:model} and \eqref{eq:stmodel} for each set of $X_\mathrm{H_2}$ and $\phi$.
Four sets of experiment conditions, with the axial bulk velocities $U_x$ of 15, 20, 25, and 30 m/s and various $X_\mathrm{H_2}$ and $\phi$, are marked with different symbols provided in legends in Fig.~\ref{fig:model_assess}b. 
Four contour lines of $U_{x,cr}^{ud} = 15$, 20, 25, and 30 m/s predicted by the present model are also plotted to provide a direct comparison in Fig.~\ref{fig:model_assess}a.
For the same inlet flow velocity, the flashback tends to occur with increasing $X_\mathrm{H_2}$ and decreasing $\phi$.
%
Since each set of symbols almost locate along the contour line of $U_{x,cr}^{ud}$ predicted by the model, the modeling results well agree with the experimental ones.

Note that the model predicts the maximum velocity of the BLF limit appears near the lean mixture for hydrogen-enriched fuels, due to the thermo-diffusive effects in the present $s_T$ model. 
The thermo-diffusive effects can significantly accelerate the propagation of turbulent flames, so the equivalence ratio with maximum $s_T$ may not agree with that for $s_L^0$. 
This makes the lean hydrogen-enriched flames more prone to the flashback.

Figure~\ref{fig:model_assess}b compares $U_{x,cr}^{ud}$ obtained from the experiment and model for the 14 cases in Table~\ref{tab:exp_cases}, with the symbols colored by the corresponding $s_L^0$.
The symbols lying close to the diagonal line demonstrates that the model gives quantitative good predictions for a range of conditions.

\section{Conclusions}\label{sec:conclusion}

We investigate the BLF of premixed hydrogen-enriched swirling flames at high pressures, using the LES-FSD method with an improved model of the local displacement speed.
To incorporate the effects of the flame stretch and heat loss, the displacement speed model employs a lookup table consisting of solutions from laminar stretched and non-adiabatic flames.

The LES-FSD result well captures the transient BLF process along the  central bluff body in the swirl burner.
The crosswind and upwind propagation modes of the rotating flame tongue observed in the experiments~\cite{Ebi2021} are reproduced in the LES-FSD.
Local distributions of the velocity and progress variable confirm that the flame mainly propagates within the boundary layer, so the flame propagation upstream is driven by the BLF.
Furthermore, the LES-FSD result accurately provides the variation of the BLF limit with the hydrogen volume fraction in fuel, via the improved model of the local displacement speed.

From the LES-FSD result, we identify the propagation mode of the BLF from the dominant propagating side of the flame tongue.
An algebraic model is then developed to predict the BLF limit of the swirling flames.
The model estimates the critical bulk velocity for given reactants and swirl number, via the balance between the flame-induced pressure rise and the adverse pressure for boundary-layer separation.
The incorporation of the propagation mode analysis and the turbulent burning velocity model~\cite{Lu2022} extends the existing model~\cite{Hoferichter2017} for non-swirling flames to swirling flames at high pressures with various fuels.
The present model is validated against 14 datasets of experiments. It well predicts the BLF limit for hydrogen volume fractions ranging from 50\% to 100\% at $p=2.5$ bar.

Note that although the heat loss effect is considered in the present LES-FSD, the simplified model on the thermal boundary conditions needs to be improved for more complex conditions with radiation and flame-wall interactions. 
Effects of the centripetal force of gas movement need to be analyzed for a more general model on the BLF limit of swirling flames.
In addition, the proposed model of the BLF limit is expected to be further validated in other experiments and practical applications.

\section*{Acknowledgement}
We gratefully acknowledge Caltech, the University of Colorado at Boulder, and Stanford University for licensing the NGA code used in this work.
Numerical simulations were carried out on the Tianhe-2A supercomputer in Guangzhou, China.
This work has been supported in part by the National Natural Science Foundation of China (Grant Nos.~91841302, 11925201, and 11988102), the National Key R\&D Program of China (No.~2020YFE0204200), and the Xplore Prize.

\appendix
\section{Methane/air swirling flames at atmospheric pressure}\label{sec:atmospheric_pressure}
\setcounter{figure}{0}
\setcounter{table}{0}
The LES-FSD for the swirl burner with a central bluff body is validated against the experimental results for the non-reacting flow and methane/air flames at atmospheric pressure.
Ebi \etal~\cite{Ebi2016,Ebi2018} reported the BLF experiments of swirling $\mathrm{CH_4}$/air flames with $p=1.0$ bar, $T_{ub}=293$ K, and $S=0.9$.
As listed in Table~\ref{tab:LES1atmCases}, there are two operating conditions with different $\phi$ and $U_x$.
The burner for the experiments is similar to that described in Section~\ref{sec:config} except for burner sizes.
The diameter and length of the combustion chamber are 100 mm and 150 mm, respectively.
The inner and outer diameters of the mixing tube are 25.4 mm and 52 mm, respectively.
The length of the mixing tube is 150 mm.

\begin{table} 	
	\caption{Operating conditions in the experiment~\cite{Ebi2016,Ebi2018} and LES-FSD of non-reacting flows and CH$_4$/air swirling flames.
    }
    \centering
	\begin{tabular}{ccccc}
		\hline
		$\mathrm{case}$	&$\phi$		&$p\;(\mathrm{bar})$ &$T_{ub}\;(\mathrm{K})$	&$U_x\;(\mathrm{m/s})$ \\
		\hline
		F1		 &0.8	&1		&293	&2.5    \\
		F2		 &1.0	&1		&293	&5.0    \\		
		\hline
	\end{tabular}
	\label{tab:LES1atmCases}
\end{table}

\begin{figure}
	\centering
	\includegraphics[width=0.5\textwidth]{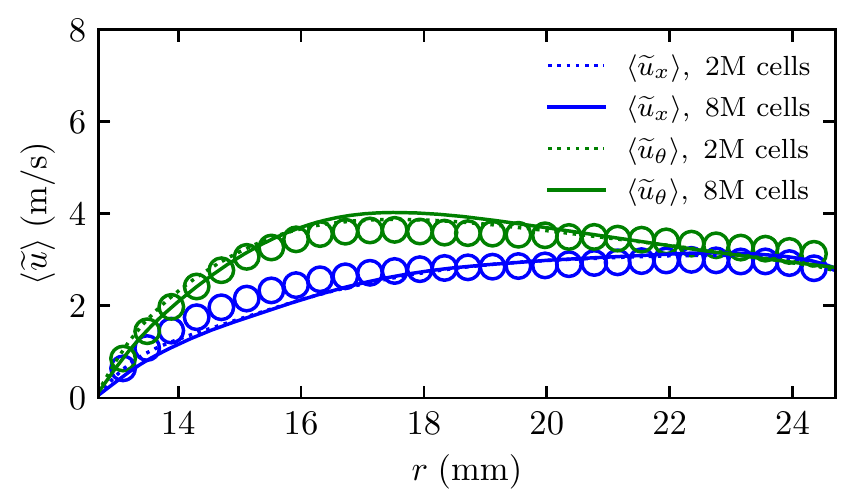}
	\caption{
        Averaged $\widetilde{u}_x$ and $\widetilde{u}_\theta$ at $x=-60$ mm obtained from les with 2 million cells (dotted line), 8 million cells (solid line), and experiments (symbols).
    }
	\label{fig:noreacting}
\end{figure}

We conducted a mesh convergence test with 2 million and 8 million cells for the non-reacting flow in case F1.
The simulation results are assessed by the velocity profiles reported in the mixing tube in the experiment~\cite{Ebi2016}.
Figure~\ref{fig:noreacting} compares the ensemble averaged velocities $\langle\widetilde{u}_x\rangle$ and $\langle\widetilde{u}_\theta\rangle$ at $x=-60$ mm obtained from the LES and experiment.
The velocity components in the LES with the two meshes and the experiment have overall good agreements.

\begin{figure}
	\centering
	\includegraphics[width=0.45\textwidth]{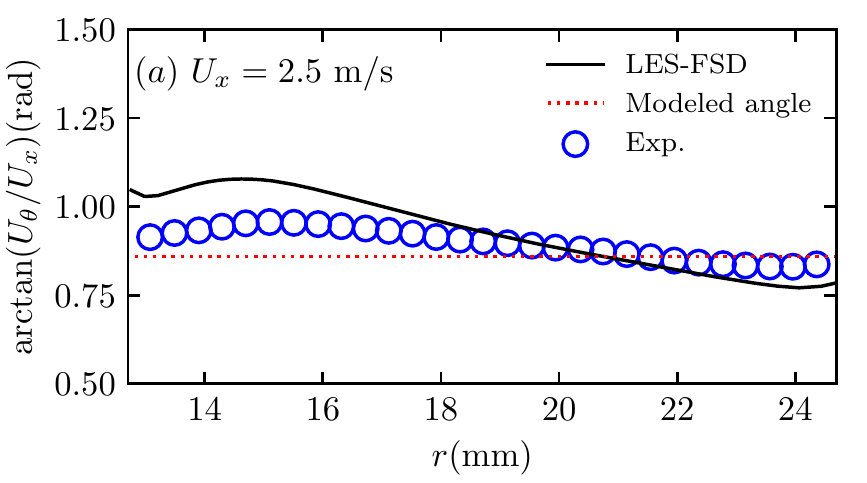}
	\includegraphics[width=0.45\textwidth]{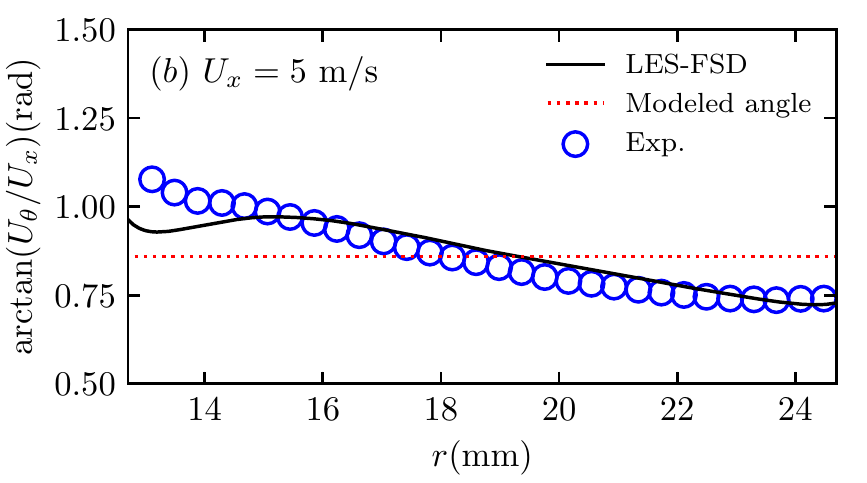}
    \caption{The angle $\alpha$ obtained by Eq.~\eqref{eq:angle}, LES, and experiments~\cite{Ebi2016} in nonreacting swirling flows of cases (a) F1 and (b) F2.}
    \label{fig:angle_model}
\end{figure}

Comparisons on the modeled $\alpha$ by Eq.~\eqref{eq:angle} and the direction of the bulk flow direction obtained in experiments~\cite{Ebi2016} and LES are presented in Fig.~\ref{fig:angle_model}.
It shows that the modeled $\alpha$ agrees well with the experimental and LES results, supporting the approximation in Eq.~\eqref{eq:angle}.

\begin{figure}
	\centering
    \includegraphics[width=\textwidth]{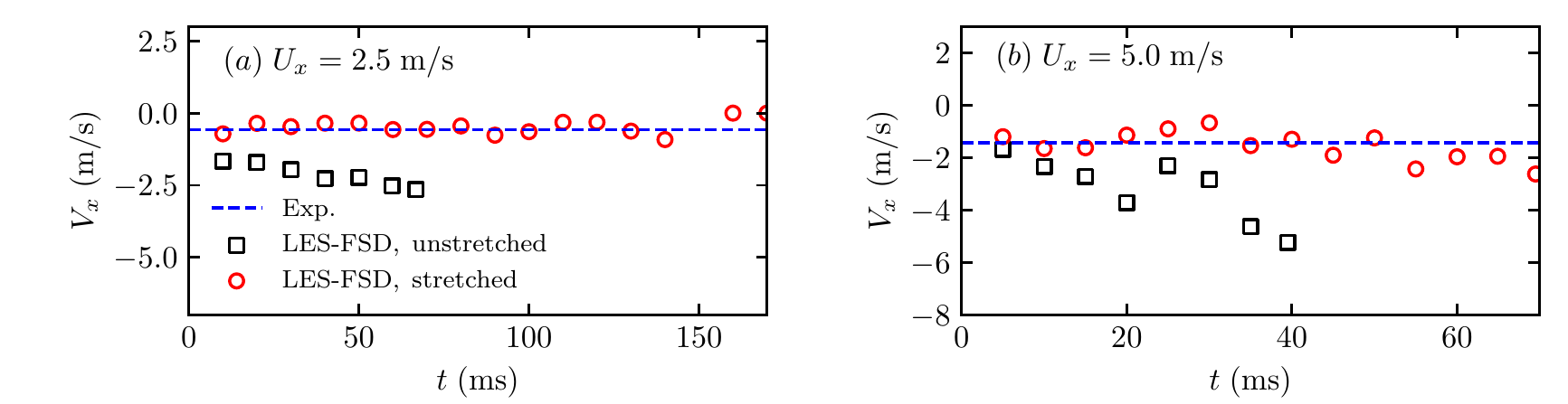}
	\caption{
        Comparisons of the absolute axial velocities of the flame tongue during BLF obtained via experiments (blue dashed lines) and LES-FSD with $\langle s_d\rangle_A$ models in Eq.~\eqref{eq:sdm1} based on unstretched flames (black squares) and Eq.~\eqref{eq:sdm2} based on stretched flames (red circles) in cases (a) F1 and (b) F2.
        }
	\label{fig:compare_sd}
\end{figure}

The models of the local displacement speed are tested for the methane/air swirling flames at atmospheric pressure.
The present model of the displacement speed model in Eq.~\eqref{eq:sdm2} is compared with a widely used model~\cite{Boger1998}
\begin{equation}\label{eq:sdm1}
	\langle s_d \rangle_{A,0}
	=
	\dfrac{\rho_{ub} s_L^0 (\widetilde{h},\widetilde{Z}) }{\overline\rho}
\end{equation}
which neglects the flame strain and flame curvature effects.
Figure~\ref{fig:compare_sd} plots the absolute axial velocity of the flame tongue during BLF in cases F1 and F2 obtained from experiments (blue dash line) and LES-FSD with $\langle s_d\rangle_A$ models in Eq.~\eqref{eq:sdm1} based on unstretched flames (black squares) and Eq.~\eqref{eq:sdm2} based on stretched flames (red circles).
The LES-FSD results are significantly improved using the present model in Eq.~\eqref{eq:sdm2}.

\bibliographystyle{elsarticle-num}
\bibliography{flashback.bib}





\end{document}